\documentclass[journal]{IEEEtran}
\usepackage{amsmath,amsfonts}
\usepackage{algorithm2e}

\usepackage{array}
\usepackage{textcomp}
\usepackage{stfloats}
\usepackage{url}
\usepackage{hyperref}
\usepackage{verbatim}
\usepackage{graphicx}
\usepackage{cite}
\usepackage{xcolor}
\usepackage{multirow}
\usepackage{booktabs}
\usepackage{soul}
\usepackage{bbm}

\usepackage{svg}
\RestyleAlgo{ruled}
\LinesNumbered
\ifCLASSOPTIONcompsoc
\usepackage[caption=false,font=normalsize,labelfont=sf,textfont=sf]{subfig}
\else
\usepackage[caption=false,font=footnotesize]{subfig}
\fi

\newcommand{\rev}[1]{\textcolor{black}{#1}}

\newcommand{\paul}[1]{\textcolor{black}{#1}}
\newcommand{\leon}[1]{\textcolor{black}{#1}}
\newcommand{\claudia}[1]{\textcolor{black}{#1}}
\usepackage{scalerel}
\usepackage{tikz}
\usetikzlibrary{svg.path}

\definecolor{orcidlogocol}{HTML}{A6CE39}
\tikzset{
  orcidlogo/.pic={
    \fill[orcidlogocol] svg{M256,128c0,70.7-57.3,128-128,128C57.3,256,0,198.7,0,128C0,57.3,57.3,0,128,0C198.7,0,256,57.3,256,128z};
    \fill[white] svg{M86.3,186.2H70.9V79.1h15.4v48.4V186.2z}
                 svg{M108.9,79.1h41.6c39.6,0,57,28.3,57,53.6c0,27.5-21.5,53.6-56.8,53.6h-41.8V79.1z M124.3,172.4h24.5c34.9,0,42.9-26.5,42.9-39.7c0-21.5-13.7-39.7-43.7-39.7h-23.7V172.4z}
                 svg{M88.7,56.8c0,5.5-4.5,10.1-10.1,10.1c-5.6,0-10.1-4.6-10.1-10.1c0-5.6,4.5-10.1,10.1-10.1C84.2,46.7,88.7,51.3,88.7,56.8z};
  }
}

\newcommand\orcidicon[1]{\href{https://orcid.org/#1}{\mbox{\scalerel*{
\begin{tikzpicture}[yscale=-1,transform shape]
\pic{orcidlogo};
\end{tikzpicture}
}{|}}}}

\usepackage{hyperref} %<--- Load after everything else
\usepackage[switch]{lineno}
\begin{document}

%\title{Scale-based Feature Generation for Seismic Data Reconstruction Improvement }
%Scale-variable Feature Generation for Seismic Reconstruction Generalization
%\title{ Generative Seismic Feature Supervision for Reconstruction Generalization}
%\title{GAN-Guided Seismic Data Reconstruction: Enhancing Supervised Learning for Improved Generalization}
\title{GAN--supervised Seismic Data Reconstruction: An Enhanced--Learning for Improved Generalization}

\author{Paul Goyes-Pe\~nafiel\textsuperscript{\orcidicon{0000-0003-3224-3747}},~\IEEEmembership{Graduate Student Member,~IEEE,} Le\'on Suarez-Rodriguez\textsuperscript{\orcidicon{0009-0004-0685-8793}}, Claudia V. Correa\textsuperscript{\orcidicon{0000-0002-1812-287X}},~\IEEEmembership{Member,~IEEE,} Henry Arguello\textsuperscript{\orcidicon{0000-0002-2202-253X}}, ~\IEEEmembership{Senior Member,~IEEE,}
        % <-this % stops a space

\thanks{Manuscript received October 23, 2023; revised December 16, 2023.}
\thanks{This work was funded by the Vicerrector\'ia de Investigaci\'on y Extensi\'on from Universidad Industrial de Santander under Project 3925.}

\thanks{P.~Goyes-Pe\~nafiel, Le\'on Suarez-Rodriguez, Claudia V. Correa and H.~Arguello are with the Department of Systems Engineering, Universidad Industrial de Santander, Bucaramanga 680002, Colombia (e-mail: ypgoype@correo.uis.edu.co; leonsuarez24@gmail.com, claudia.correa@correo.uis.edu.co, henarfu@uis.edu.co)}

}% <-this % stops a space}

\markboth{IEEE Transactions on Geoscience and Remote Sensing,~Vol.~00, No.~0,~2024}%
{Shell \MakeLowercase{\textit{et al.}}: A Sample Article Using IEEEtran.cls for IEEE Journals}


\maketitle

\begin{abstract}
Seismic data interpolation of \rev{irregularly missing traces} plays a crucial role in subsurface imaging, enabling accurate analysis and interpretation throughout the seismic processing workflow. Despite the widespread exploration of deep supervised learning methods for seismic data reconstruction, several challenges remain open. Particularly, the requirement of extensive training data and poor domain generalization due to the seismic survey's variability poses significant issues. To overcome these limitations, this paper introduces a deep-learning-based seismic data reconstruction approach that leverages data redundancy. This method involves a two-stage training process. First, an adversarial generative network (GAN) is trained using synthetic seismic data, enabling the extraction and learning of their primary and local seismic characteristics. Second, a reconstruction network is trained with synthetic data generated by the GAN, which dynamically adjusts the noise and distortion level at each epoch to promote feature diversity. This approach enhances the generalization capabilities of the reconstruction network by allowing control over the generation of seismic patterns from the latent space of the GAN, thereby reducing the dependency on large seismic databases. Experimental results on field and synthetic seismic datasets both pre--stack and post--stack show that the proposed method outperforms the baseline supervised learning and unsupervised approaches such as deep seismic prior and internal learning, by up to 8 dB of PSNR.
\end{abstract}


\section{Introduction}

\IEEEPARstart{S}{eismic} data reconstruction plays a crucial role in geophysics, enabling accurate analysis and interpretation for various applications including oil and gas exploration, earthquake monitoring, and subsurface imaging. Reliable seismic data reconstruction is essential for obtaining high-fidelity subsurface images and making informed decisions in geophysical studies. 

In recent years, deep learning-based (DL) methods have shown great potential for \rev{irregularly missing trace} interpolation, offering the ability to fill in seismic data gaps and recover missing information. These approaches often rely on supervised learning, where a reconstruction network is trained using pairs of sub-sampled and complete seismic data \cite{Fang2021,Park2019,Wang2019,Wang2020,Yeeh2020}. While fully-supervised DL can yield accurate results, it comes with a significant caveat - the effectiveness of the network is highly dependent on the quality and diversity of the training dataset. Therefore, supervised models have poor generalization capabilities, which limits their performance on new testing data exhibiting different structures compared to those from the training data \cite{fernandez2022}. For instance, supervised DL models for seismic trace reconstruction are typically trained for a particular survey, employing a portion of the dataset for training, and the remaining data is used for testing \cite{Wang2019, Yu2022, He2022, Mandelli2018a}. To improve the network performance in the testing data, some seismic applications have used transfer learning from synthetic to field seismic data \cite{Wang2020}. Alternatively, multiple datasets have been used in the training, thereby avoiding the need to retrain the models for each new survey \cite{Dou2023}. 
On the other hand, unsupervised and self-supervised learning approaches, such as deep image prior or internal learning, have been explored to train reconstruction networks without using external datasets, as these methods only rely on incomplete measured data to reconstruct the complete seismic data \cite{Goyes-Penafiel2023,Kong2022,Liu2021}. However, these models still have a limited generalization capacity as they require training for each particular dataset. Additionally, it is worth pointing out that hyperparameter configurations vary with the dataset, implying extension computations for parameter tuning \cite{Rodriguez-Lopez2023,Kong2020, Wang2020a, Goyes-Penafiel2023}. The challenges mentioned above for supervised and unsupervised learning prevent current deep learning approaches from capturing the complexity of subsurface structures and accommodating differences in data acquisition, which in turn limits their generalization capabilities across various datasets.

This work proposes an approach to address the generalization reconstruction issue without using external field training datasets. Thus, we overcome the pervasive issue of insufficient training data within seismic acquisition in the petroleum industry due to data ownership (proprietary or non-exclusive data).
Our novel approach exploits the power of Generative Adversarial Networks (GANs) to supervise traditional reconstruction learning. Specifically, GANs are renowned for their ability to generate highly diverse data by learning intricate patterns from training data \cite{Wei2021a,Siahkoohi2018a, Wei2021 ,Chang2021}, and they have shown their ability to generate seismic data in different scenarios, maintaining coherence between structures and seismic signals \cite{Dou2023,Song2022,Kaur2019,Chang2019}. In the proposed approach, a GAN is trained to generate synthetic seismic images exhibiting a wide range of \paul{seismic events} with varying complexities \paul{related to the dips and curvature of the reflectors}. These synthetic images are then utilized within a reconstruction neural network so that the recovery process is supervised by the synthetic data generated by the GAN. Further, we present a strategy for controlling the variance of the latent space distribution during the training of the reconstruction network. This strategy aims to achieve diverse image generation, emulating seismic scenarios with low-quality data, \paul{such as limited resolution, noise, and artifacts}. Our approach delivers high-quality experimental results, thus demonstrating its effectiveness and generalization capabilities when applied to a diverse range of seismic datasets, providing a robust and adaptable solution for seismic data reconstruction. In particular, the proposed method outperforms state-of-the-art supervised and unsupervised learning approaches, such as deep seismic prior and internal learning, in up to 8 dB of PSNR.

\section{GAN-Guided Seismic reconstruction}
\rev{Irregularly} missing traces in seismic data are typically caused by environmental or equipment constraints. Seismic data with missing traces can be mathematically modeled %through a masking operator $\boldsymbol{\Phi}$ that maps a subset of traces to zero-valued columns 
as
%The missing seismic data reconstruction can be modeled through a subsampling operator $\boldsymbol{\Phi}$ simulating the missing traces caused by environmental or equipment constraints. Thus, mathematically the problem is given by
\begin{equation}
\label{eq:acq}
    \mathbf{Y}= \boldsymbol{\Phi} \odot \mathbf{X},
\end{equation}
where $\mathbf{Y},\mathbf{X} \in \mathbb{R}^{M\times N}$ are the corrupted/incomplete and the complete seismic data, respectively, with $M$ denoting the time samples and $N$ the number of traces in the full data; $\odot$ is the Hadarmard product; $\boldsymbol{\Phi} \in \{0,1 \}^{M\times N}$ is a masking operator that maps a subset of traces to zero-valued columns in $\mathbf{Y}$. The masking operator is modeled as $ \mathbbm{1}_{N} \otimes \mathbf{s}^{T}$, where $\mathbf{s}$ is a binary indicator vector that selects the measured traces; $\otimes$ is the Kronecker product \cite{kronProd} and $\mathbbm{1}_N$ is a $N$-long one-valued vector.

While the structure of seismic data representation $\mathbf{X}$ can vary depending on factors such as geometry (e.g., split-spread or inline offset \cite{Galvis2020, Rodriguez-Lopez2023}, cross-spread \cite{Villarreal2019,Goyes-Penafiel2021}, swath \cite{Goyes-Penafiel2023a}); sorting (e.g., common-shot-gather, common-receiver-gather, common-midpoint); and processing (pre-stack or post-stack), the underlying similarity and coherence of the structures persist \cite{Yilmaz2001a,Cordsen2000,Chaouch2006,Liner2016a}. This enduring coherence arises from the fact that the acquisition process predominantly captures both, the signal and coherent noise stemming from subsurface wave propagation \cite{Goyes-Penafiel2023a}. The redundant structures in seismic images have been exploited in seismic reconstruction and denoising using local information represented by patches \cite{Yang2022a,Liu2011,Saad2023,Wang2021,Claerbout2014,Chen2019}. %, \paul{further additional details about patching techniques are described in \cite{Claerbout2014GeophysicalExample,Chen2019ImprovingLearning}.}

\begin{figure}[b!]
    \centering
    \includegraphics[width=0.7\columnwidth]{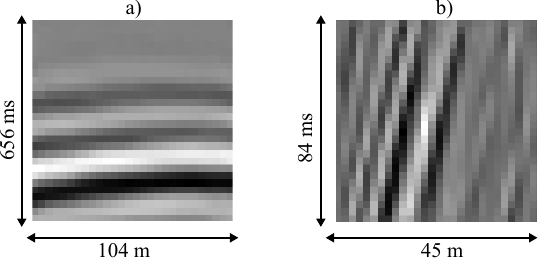}
    \caption{Two $28 \times 28$ patches showing typical seismic events in a different space-time length extracted from (a) $128 \times 128$, and (b) $1501 \times 96$ common-shot-gathers.} 
    \label{fig:patch}
\end{figure}
However, these recurrent structures are highly sensitive to the influence of varying scales within seismic images. More specifically, since the number of samples in the time domain ($M$) greatly exceeds those in the spatial domain ($N$), local seismic structures at different image scales exhibit slope and shape distortions. This effect is depicted in Fig.~\ref{fig:patch}, which analyzes the distortion over a single reflector in  $28\times28$ patches extracted from two different common-shot-gathers, i.e., a $128 \times 128$ and a $1501 \times 96$ image. Although the seismic structures are identifiable in both cases, it can be clearly seen that employing the same patch size at different image scales causes a change in time and space sampling rates. This phenomenon restricts deep learning-based reconstruction algorithms based on convolutional architectures to adopt inputs of fixed dimensions (e.g. deep image prior-based \cite{Rodriguez-Lopez2023}). Therefore, supervised training approaches are highly contingent upon the dataset they are trained on, hindering their generalization capabilities. 

To overcome these limitations of deep learning-based seismic reconstruction methods, this work proposes to learn the representation of local signal structures and scale-variable seismic patterns through a generative model, which enhances the generalization capability by feeding the reconstruction network with diverse images as it will be explained in the following subsections.

\subsection{Generative Seismic Patches} 
%To avoid dependencies on real training data, we
The first stage of the proposed reconstruction method consists of generating synthetic $28 \times 28$ patches of high variability. These patches simulate the local structures within seismic images by harnessing their inherent repetitiveness and adaptability across different scales \cite{Lai2016,Ou2023}, thus avoiding dependencies on field datasets for training. To achieve this, we employ the well-known Wasserstein Generative Adversarial Network (WGAN) with Gradient Penalty (GP) following the training strategy proposed by \cite{Gulrajani2017}. The loss function of the WGAN--GP exhibits convergence properties related to the quality of the generated images, which is advantageous during training, as it eliminates the need for manual monitoring of training stability and the risk of converging into specific features \cite{Arjovsky2017,Gulrajani2017}. In this work, the WGAN-GP is trained only using the synthetic dataset from \cite{Goyes-Penafiel2023}, which contains rectangular and distorted seismic data of size $128 \times128$, $750\times 100$, and $128\times100$.

\begin{figure}[t]
     \centering
     \includegraphics[width=.7\columnwidth]{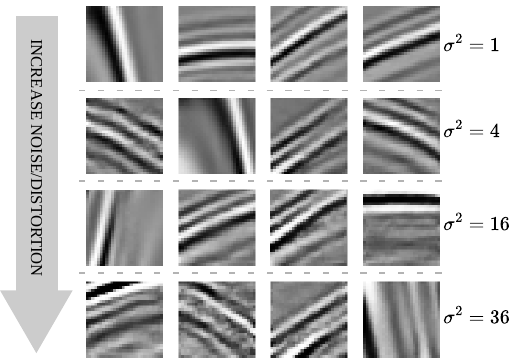}
 \caption{Local seismic structures represented by patches generated with \paul{different variance values}  $\sigma^2=\{1,4,16,36 \}$.% Blue highlighted patches illustrate \paul{ examples of linear events} distorted by high values of $\sigma^2$.
 }
     \label{fig:features}
 \end{figure}

The generative neural model employs a generator network $\mathcal{G}$ that learns the probability distribution of the synthetic training data samples, given a fixed \claudia{standard normal} distribution of the latent vectors $\mathbf{z}\sim \mathcal{N}(0,1)$ \cite{Van2020,Asperti2023}, so that the trained model can synthesize realistic images as $\mathbf{X} = \mathcal{G}(\mathbf{z})$. A key aspect of the proposed generative seismic patches lies in that during inference, we query the network to provide patch samples from probability distributions with different variances. \paul{Specifically, 
 a latent vector $\mathbf{z}_{\sigma}$ follows a multivariate normal distribution with mean $0$ and covariance matrix $\sigma^2 \boldsymbol{I}$, denoted as} $\mathbf{z}_{\sigma}\sim \mathcal{N}(0,\sigma^2 \boldsymbol{I})$, \paul{where $\sigma^2$ is the variance and $\boldsymbol{I}$ is an identity matrix}. Thus, we effectively introduce more variability and randomness into the latent space. \paul{Consequently, the generator is less constrained and more prone to producing images divergent from the data manifold on which the GAN was trained, thereby resulting in noisy and distorted images that may not align with the ideal synthetic seismic patches.} These distortions, \rev{however, are able to} mimic  low-quality seismic features. For instance, Fig.~\ref{fig:features} illustrates multiple synthetic patches generated by the proposed model, using different \claudia{variance values} $\sigma^2$, where the blue highlighted patches depict \paul{linear events} that are progressively distorted as $\sigma^2$ increases, simulating noisier seismic patterns.

The \paul{trained generator $ \mathcal{G}$ is }later used to guide the training of the reconstruction \paul{task}, as described in the following section.

\subsection{GAN-supervised Reconstruction Learning}
\begin{figure}[t!]
    \centering
    \includegraphics[width=1\columnwidth]{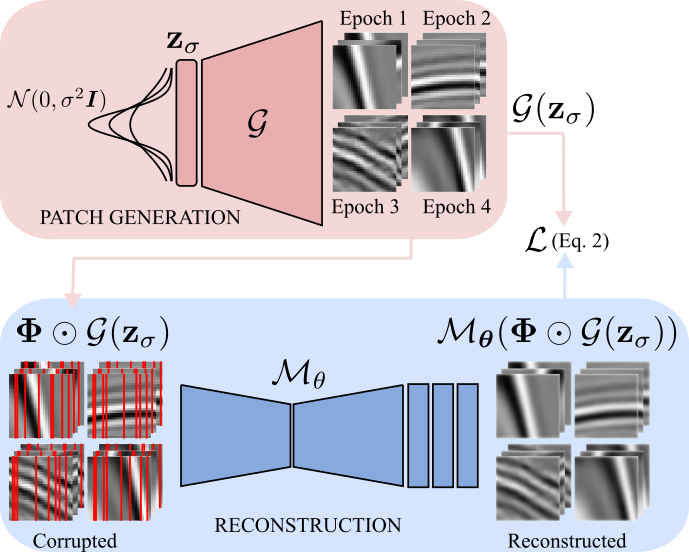}
    \caption{GAN-guided seismic reconstruction training scheme illustrating a batch size $B=3$ and 4 epochs. Red vertical lines represent the missing traces. The masking pattern and $\sigma \in [1,3]$ are randomly fixed at every epoch. }
    \label{fig:scheme}
\end{figure}

A GAN-supervised learning approach is proposed here, aiming to guide and enhance the reconstruction network's generalization capabilities. Specifically, the GAN model $\mathcal{G}$ is used to generate a batch of $B$ seismic patches $\mathcal{G}(\mathbf{z}_\sigma)$ at each training epoch of the reconstruction network $\mathcal{M}_{\boldsymbol{\theta}}$,  \paul{with trainable parameters $\boldsymbol{\theta}$}. Thus, $\mathcal{M}_{\boldsymbol{\theta}}$ is exposed to a continuously changing set of subsampled input patches as shown in Fig.~\ref{fig:scheme}. The key advantage of this approach is the introduction of diversity and adaptability into the training process. Additionally, during the reconstruction learning, we employ latent vectors $\mathbf{z}_\sigma$ that are far from what the generator $\mathcal{G}$ was trained on (i.e., outside the typical confidence interval), which may result in out-of-distribution data patches, leading to noisy or distorted generated images simulating degradation and variable-scale seismic patterns as is shown in Fig.~\ref{fig:features}.

%To leverage the internal features of the generated dataset $\mathcal{D}(\mathbf{z}_\sigma)$, 
\subsubsection{Reconstruction Loss Function}
\paul{We employ the hybrid reconstruction loss function proposed by \cite{Yu2022}, which combines the Mean Absolute Error (MAE) and the Structural Similarity Index (SSIM) \cite{1284395} metrics, and is expressed as}
\begin{equation}
\label{eq:loss}
\begin{split}
    \mathcal{L}= \frac{1}{B}\left \| \mathcal{G}(\mathbf{z}_\sigma) - \mathcal{M}_{\boldsymbol{\theta}}(\boldsymbol{\Phi} \odot \mathcal{G}(\mathbf{z}_\sigma)) \right \|_1 + \\
    1 - \mathrm{SSIM}[\mathcal{G}(\mathbf{z}_\sigma), \mathcal{M}_{\boldsymbol{\theta}}(\boldsymbol{\Phi} \odot \mathcal{G}(\mathbf{z}_\sigma))],
\end{split}
\end{equation}
\paul{where the first term %of the equation 
computes the MAE between the generated seismic patches and the reconstruction of the subsampled generated seismic patches, with $B$ as the batch size. The second term aims to maximize the SSIM between the generated seismic patches and the reconstruction of the subsampled generated seismic patches, with a maximum value of one. The masking operator that simulates a corrupted image is denoted by $\boldsymbol{\Phi}$, as in Eq.~\eqref{eq:acq}. This loss function balances pixel-level accuracy and perceptual image quality, considering luminance, contrast, and structure \cite{Li2022,Yu2022,Jin2023}.} % with missing traces at random positions with 15\%, 25\%, and 35\% compression ratios. 

\begin{figure*}[ht]
\centering
\includegraphics[width=0.8\textwidth]{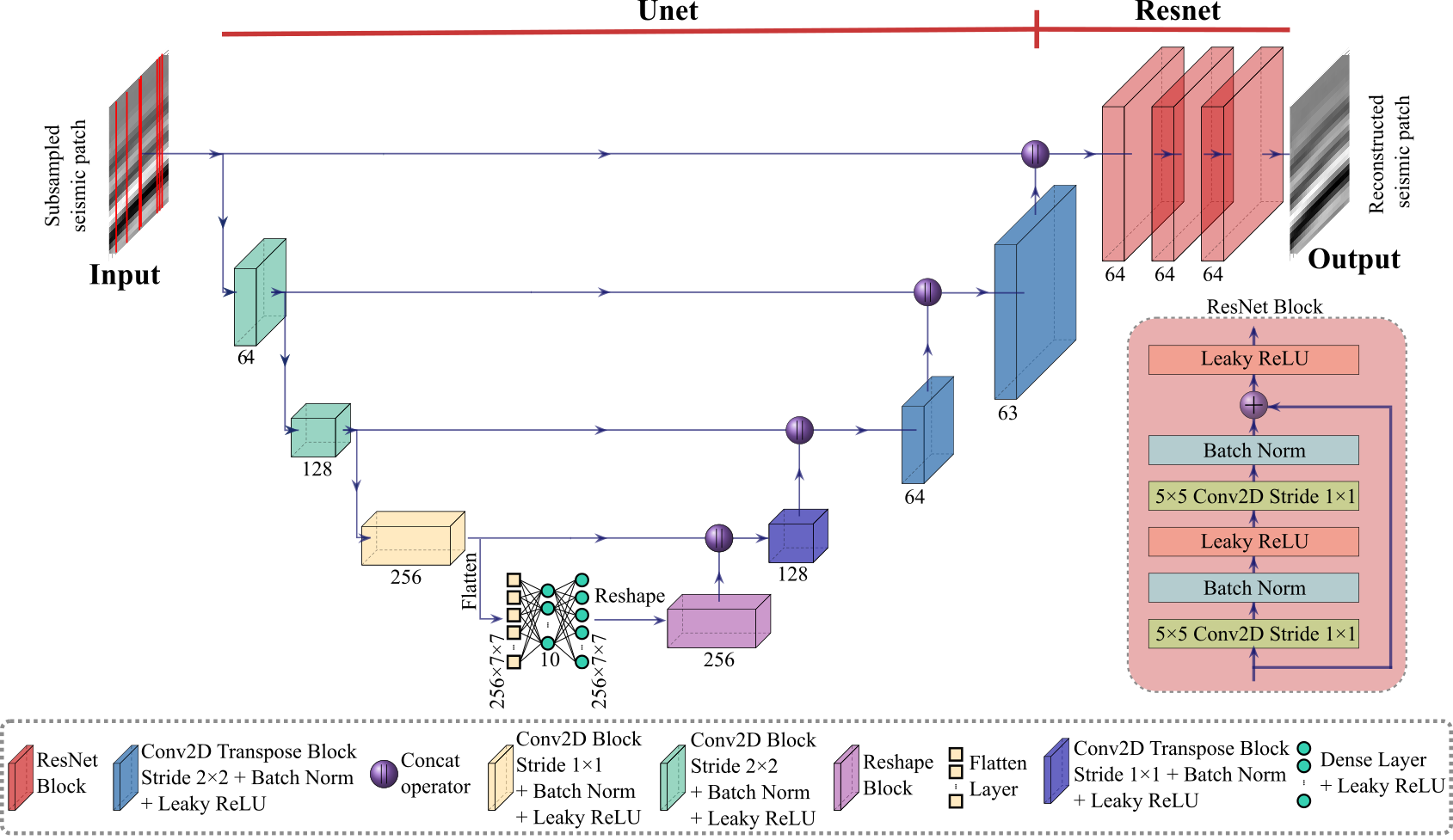}
\caption{Architecture of the reconstruction neural network $\mathcal{M}_{\boldsymbol{\theta}}$ composed of a Unet \cite{UNET} connected to three Resnet blocks \cite{resnet_he}.The bottleneck of the Unet corresponds to a \textit{fully-connected} neural network with three layers of 12544 \paul{(from flattening a $256 \times 7 \times 7$ conv layer)}, 10, and 12544 neurons, respectively.}
\label{fig_1:network}
\end{figure*}

The GAN-supervised learning can be expressed as the minimization of Eq.~\eqref{eq:loss} given by 
\begin{equation}
\label{eq:opt}
    \underset{\boldsymbol{\theta} }{\mathrm{\min}} \left \{ \mathcal{L}(\boldsymbol{\theta};\mathcal{G}(\mathbf{z}_\sigma))+\left \| \boldsymbol{\theta} \right \|_1 \right \},
\end{equation}
\paul{where the $\ell_1$ regularization over $\boldsymbol{\theta}$ prevents  overfitting and avoids the trainable parameters weigh heavily \cite{Chen2019}.}

From Eq.~\eqref{eq:opt}, it can be noted that the proposed GAN-supervised learning does not require external field data for training $\mathcal{M}_{\boldsymbol{\theta}}$. Consequently, \rev{the proposed supervised learning relies solely on the outputs of the GAN model, which in the context of this paper was exclusively trained using synthetic data.}

The trained $\mathcal{M}_{\boldsymbol{\theta}}$ is then used to reconstruct missing seismic traces within any test seismic image $\mathbf{Y}$. Specifically, the operator $\wp$ extracts $28\times28$ patches from $\mathbf{Y}$ with a stride of $1\times1$, and re-arranges them as $k\times 28\times28\times1$ structures (patches), where $k$ denotes the number of patches to be processed by $\mathcal{M}_{\boldsymbol{\theta}}$. Hence, the patches with the recovered traces are used to construct the non-corrupted seismic image $\hat{\mathbf{X}}$ by applying the \paul{inverse patching} operator $\wp^{-1}$ as follows

\begin{equation}
\label{eq:recon}
    \hat{\mathbf{X}}=\wp^{-1}(\mathcal{M}_{\boldsymbol{\theta}}(\wp (\mathbf{Y}))),
\end{equation}
%where $\mathbf{X}$ is the non-corrupted version of $\mathbf{Y}$.
for the overlapping pixels within patches, we employ the median operation to effectively address any edge-related concerns. Thus, Eq.~\eqref{eq:recon} is the proposed solution to the problem in Eq.~\eqref{eq:acq}. We remark that the training of $\mathcal{M}_{\boldsymbol{\theta}}$ \leon{relies solely on the WGAN-generated images and}, does not employ information or knowledge about $\mathbf{Y}$.

\vspace{0.3cm}
\subsubsection{Neural Network Configuration}

The proposed network $\mathcal{M}_{ 
\boldsymbol{\theta}}$ combines a 2D Unet \cite{UNET} with Resnet blocks  \cite{resnet_he} to extract the spatial features from the seismic patches. Fig.~\ref{fig_1:network} shows a schematic representation of the employed network architecture. The downsampling and upsampling operations are addressed by standard convolution and transpose convolution layers with $2 \times 2$ strides, respectively. Skip connections were included between the encoder and the decoder to facilitate the fusion of low-level \paul{(edges, corners, or blobs)} and high-level \paul{(structural patterns and events)} features. Before the Resnet blocks, we concatenated the input and the output of the Unet to enhance the spatial information. To ensure consistency across all layers, batch normalization with a momentum of $0.5$ and the Leaky ReLU activation function are employed throughout. The three Resnet blocks in the image reconstruction process gradually enhance the patch quality before reaching a convolutional layer with a single filter and a Sigmoid activation function, which generates the seismic patch with the reconstructed traces. It is worth noting that although previous works have employed similar neural network architectures for data reconstruction, denoising, and segmentation \cite{Zhang2022,8309343,Venkatesh}, this work aims at providing a powerful guided learning approach for reconstruction tasks, rather than proposing new or sophisticated neural network architectures.

%%% DON'T DELETE IT !!
%\leon{The Unet is divided into: (I) \textit{\textbf{An encoder}} for low-level feature extraction, capturing simple patterns like edges, corners, or blobs. It compresses the input data into a lower-dimensional latent representation by employing $2\times 2$ stride convolutions. (II) \textit{\textbf{A bottleneck}} with a flattened--layer that collapses the spatial dimensions of the preceding block's output, reducing it from $7\times 7\times 256$ to $12,544$. It is followed by two dense layers and a reshape operation that transforms the output from the last dense layer into a $7\times 7\times 256$ volume. (III) \textit{\textbf{The decoder}} captures high-level features such as structural patterns and events, the decoder decompresses the latent representation into a volume of $28\times28\times63$ rather than directly reconstructing the image. To enhance the spatial information before the  Resnet blocks, we concatenate the input image with the decoder's output, resulting in an input volume of $28\times 28\times 64$ for the Resnet.}
%The strides, filters, and number of parameters are summarized in Table \ref{table_1:network_layer_details} for each layer in the Unet.
% capturan patrones simple, ejemplo bordes, high-level: patrones mas complejos como texturas y formas

\section{Experiments and Results}
\label{sec:experiments}
\textbf{Training details:} 
The WGAN-GP model was trained for 1000 epochs following the implementation in \cite{KumarKeras}, using the synthetic dataset from \cite{Goyes-Penafiel2023}. On the other hand, the reconstruction network $\mathcal{M}_{\boldsymbol{\theta}}$ was trained for 2000 epochs using the Adam optimizer \cite{KingBa15}. Each training epoch employed a batch of $B=500$ synthetic patches generated by the GAN model. To ensure proper convergence of the learning process, we set the learning rate at $10^{-3}$ with 0.6 exponential decay every 450 epochs. The neural models were implemented with the Keras API and TensorFlow 2.x. Further implementation details can be found in the project's repository.

\begin{figure}[ht]
    \centering
    \includegraphics[width=\columnwidth]{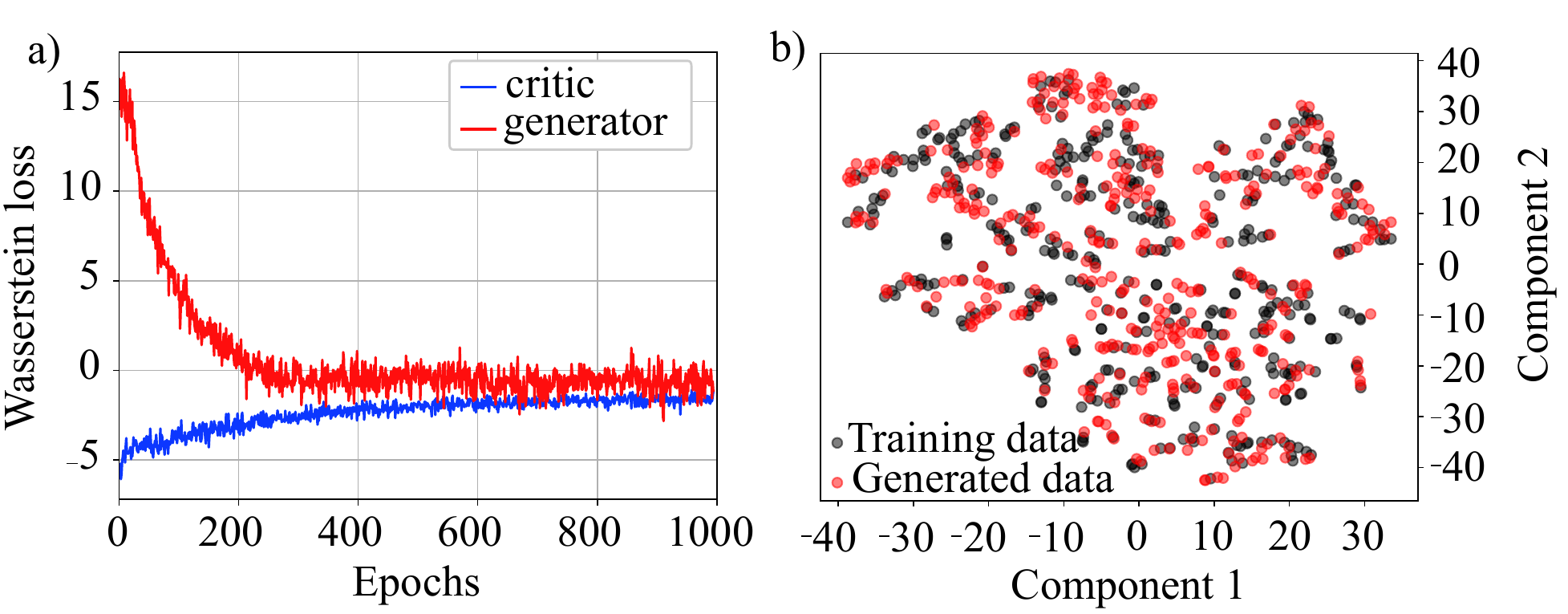}
    \caption{ \paul{a) Generator and critic loss of the WGAN. b) 2-dimensional t-SNE scatter plot of the generated (red) and original images from the training data (black).}}
    \label{fig:distributions}
\end{figure}

\textbf{Patch generator evaluation:} \paul{To analyze the fidelity of the generated images 
concerning the training data set, we first analyze the convergence curves of the loss function of the WGAN throughout training, as suggested in \cite{WGAN_ORIGINAL}. The generator loss function of the WGAN, depicted in Fig. \ref{fig:distributions}a, exhibits convergence properties, indicating a correlation between lower errors and higher quality of the generated images \cite{WGAN_ORIGINAL}. Similarly, the minimized critic loss suggests that the critic assigns similar scores to both, the generated and training data, indicating that the generator produces samples with similar characteristics to the training data, thus reducing the Wasserstein's distance between the two distributions.} %Fig. \ref{fig:distributions}a illustrates the loss convergence of the critic and generator networks, demonstrating the high quality of the generated images. Furthermore, (II) 

\paul{Furthermore, we compared the training and generated data distributions employing the scatter plot in Fig. \ref{fig:distributions}b. Specifically, we %performed dimensionality reduction on both the training data and the generated images 
projected both data sets to a two-dimensional space using the t-distributed Stochastic Neighbor Embedding (t-SNE) \cite{tsne_paper}. As it can be noted in Fig. \ref{fig:distributions}b both distributions overlap in the scatter plot. This indicates that the WGAN can generate \textit{realistic} images that closely resemble those in the training dataset.}

The effectiveness of the proposed method was evaluated through three different experiments involving the reconstruction of missing traces using field and synthetic data in both, pre-stack and post-stack scenarios, as well as 2D shot-gather reconstruction. All experiments were conducted utilizing an NVIDIA Tesla P100 16-GB GPU. To quantitatively evaluate the quality of the reconstructed data, we employed the Peak Signal-to-Noise Ratio (PSNR) and the Structural Similarity Index (SSIM) metrics following \cite{Goyes-Penafiel2023}.

\subsection{Experiment I}

\begin{figure*}[ht]
    \centering
    \includegraphics[width = \textwidth]{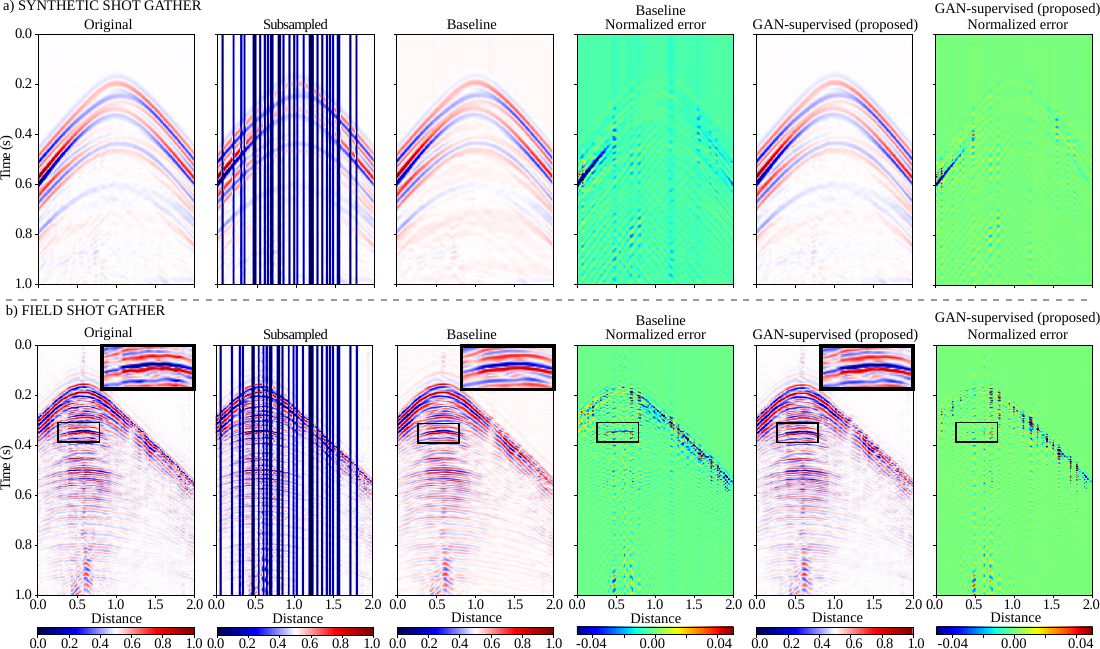}
    \caption{Comparison of pre-stack data with the reconstructed traces from Experiment I using a) the synthetic and b) the field dataset (Stratton survey). \rev{The normalized error images highlight the main differences between the proposed and baseline methods.} }
    \label{fig:res1}
\end{figure*}

In this first experiment, the proposed method was compared with a classical training scheme in a subsampling scenario with $30\%$ missing traces from a single shot-gather. The baseline model uses the network architecture $\mathcal{M}_{\theta}$ from Fig.~\ref{fig_1:network}, but instead of employing the synthetic patches generated by the GAN as the training set, it uses the dataset on which the GAN was originally trained (as described in section IIA). Both models were trained with the same training configuration parameters from in section II.B.2. This experiment considers two different test data sets: the first one comprises a synthetic shot-gather with dimensions $M=1000$ and $N=101$; the second dataset is a field shot-gather from the Stratton survey \cite{Stratton_3D_survey} with dimensions $M=700$ and $N=100$. 

\begin{figure}[ht]
    \centering
    \includegraphics[width = .85\columnwidth]{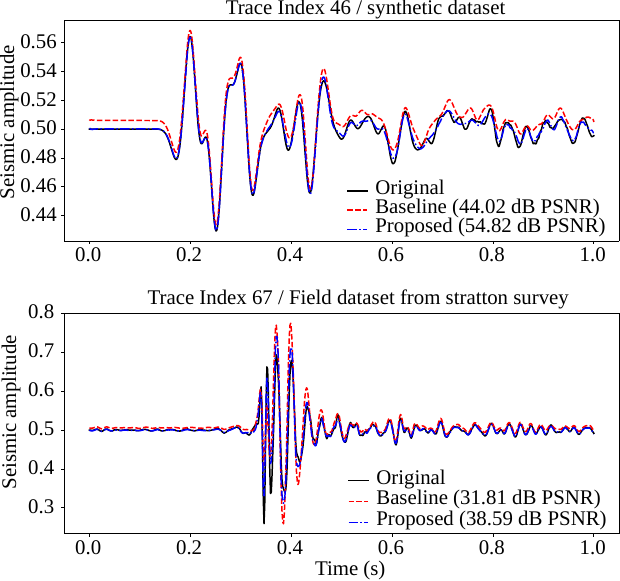}
    \caption{Visual comparison of a recovered trace using the baseline and the proposed method for Experiment I: (Top) Trace 46 of the synthetic data and, (Bottom) Trace 67 of the field data from the Stratton survey \cite{Stratton_3D_survey}.}
    \label{fig:tracecpm}
\end{figure}

\begin{table}[ht]
\centering
\caption{Average reconstruction PSNR and SSIM for recovered traces in experiment I.}
\label{table:exp1}
\begin{tabular}{llcc}
\hline
Dataset                    & Model    & SSIM  & PSNR (dB)       \\ \hline
\multirow{2}{*}{Synthetic \cite{Goyes-Penafiel2023}} & Baseline & 0.993 & 41.32          \\ \cline{2-4} 
                           & Proposed      & \textbf{0.996} & \textbf{47.92} \\ \hline
\multirow{2}{*}{Stratton \cite{Stratton_3D_survey}}     & Baseline & 0.956 & 32.30          \\ \cline{2-4} 
                           & Proposed      & \textbf{0.968} & \textbf{33.68} \\ \hline
\end{tabular}
\end{table}
Table~\ref{table:exp1} summarizes the reconstruction metrics for Experiment I, where it can be noted that the proposed method outperforms the baseline for both data sets. Specifically, the proposed method yields superior results in amplitude estimation, with improvements of 6.6 dB and 1.4 dB of PSNR for the synthetic and field data, respectively. The refinement in amplitude reconstruction is visually evident in Fig.~\ref{fig:tracecpm}, where the red line depicts the reconstructed seismic trace from the baseline, and the blue line is the recovered trace from the proposed method, which are compared to the original trace in black. The poor baseline estimation leads to a phase distortion effect in the amplitudes, significantly reducing the individual PSNR of the reconstructed trace.

Fig.~\ref{fig:res1} presents the reconstructed traces within the shot-gather along with the normalized error images. These results show the distribution of the major differences in amplitude estimation, with the proposed method exhibiting lower errors concentrated in less complex seismic features, as indicated by the arrows.

\subsection{Experiment II}
The second experiment employs the Alaska post-stack dataset \cite{Alaska_central_1981} with $M=640$ and $N=384$, focused on the deepest part of the original post-stack image, to account for a noisy field reconstruction scenario with $35\%$ missing traces. The proposed method was compared against unsupervised learning approaches, i.e., Deep Seismic Prior (DSP) \cite{Liu2021}, and Internal Learning (IL) \cite{Wang2020a}.

\begin{figure*}[ht]    
    \centering
    \includegraphics[width=0.9\textwidth]{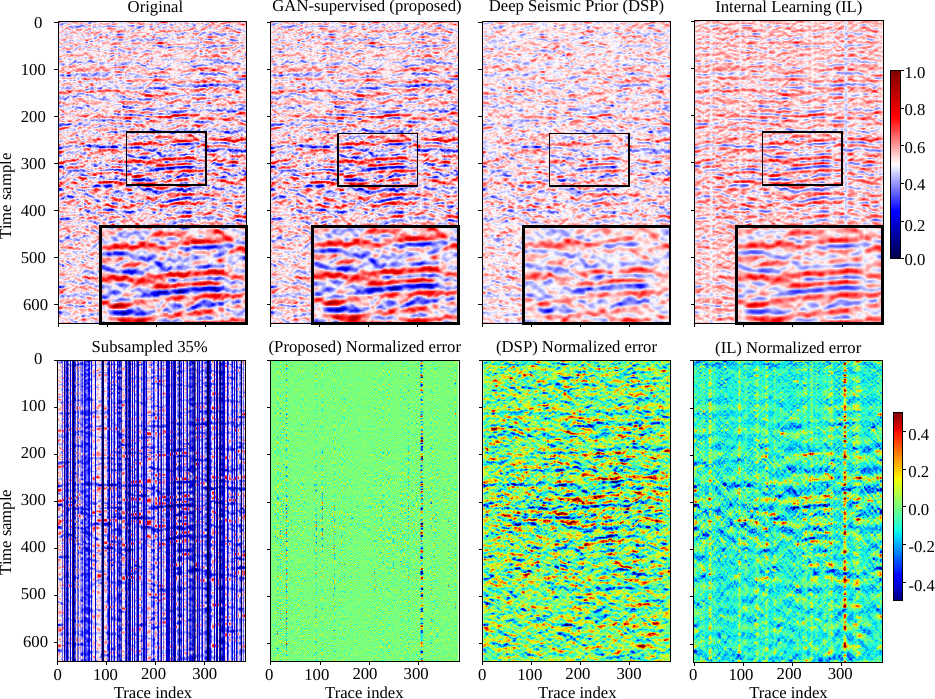}
    \caption{Reconstructions of $35\%$ \rev{irregular and random} missing traces from field Alaska dataset in Experiment II, using IL~\cite{Liu2021}, DSP~\cite{Wang2020a}, and the proposed method. }
    \label{fig:exp2}
\end{figure*}

Fig.~\ref{fig:exp2} compares the obtained results for all methods. It can be seen that the highlighted seismic events at time samples 300 and 400 were accurately recovered by the proposed technique with minimal normalized errors. Conversely, the DSP and IL methods overestimated seismic amplitudes, resulting in significant errors. In particular, the IL method generated artifacts in the reconstructed image.

\begin{table}[ht]
\centering
\caption{Average PSNR and SSIM metrics for the reconstructed traces of Alaska field dataset in Experiment II.}
\label{table:exp2}

\begin{tabular}{ccc}
\hline
Method    & SSIM                   & PSNR (dB)               \\ \hline
Proposed & \textbf{0.869 $\pm$ 0.052} & \textbf{31.49 $\pm$ 1.659} \\ \hline
DSP~\cite{Wang2020a}     & 0.604 $\pm$ 0.048          & 23.72 $\pm$ 0.652          \\ \hline
IL~\cite{Liu2021}       & 0.806 $\pm$ 0.075          & 26.29 $\pm$ 1.277          \\ \hline
\end{tabular}
\end{table} 

It is noteworthy that while both the IL and DSP methods utilized measured traces from the same shot-gather to train the reconstruction network, the proposed method did not employ any data from the field dataset. Still, it can provide accurate reconstructions of the missing traces. The summary metrics listed in Table~\ref{table:exp2} corroborate the visual outcomes, with the proposed method exhibiting reliable reconstructions that surpassed those from the IL and DSP methods by up to 8 dB of PSNR and 0.2 SSIM.

\subsection{Experiment III}
Experiment III involves the reconstruction of shot-gathers, which is more challenging than trace reconstruction.  \rev{Since the proposed method is based on 2D reconstruction as described in Eq.~\eqref{eq:acq}, the reconstruction of the 3D data for this experiment was carried out in a slice-wise fashion in the direction of the receivers, i.e., receiver-gather.}  The dataset used in this experiment is a marine acquisition known as Mobil AVO viking graben \cite{avo}, from which we extracted a subset of $256$ samples in time, $64$ receivers, and $28$ shots.

\begin{figure}[ht]    
    \centering
    \includegraphics[width=\columnwidth]{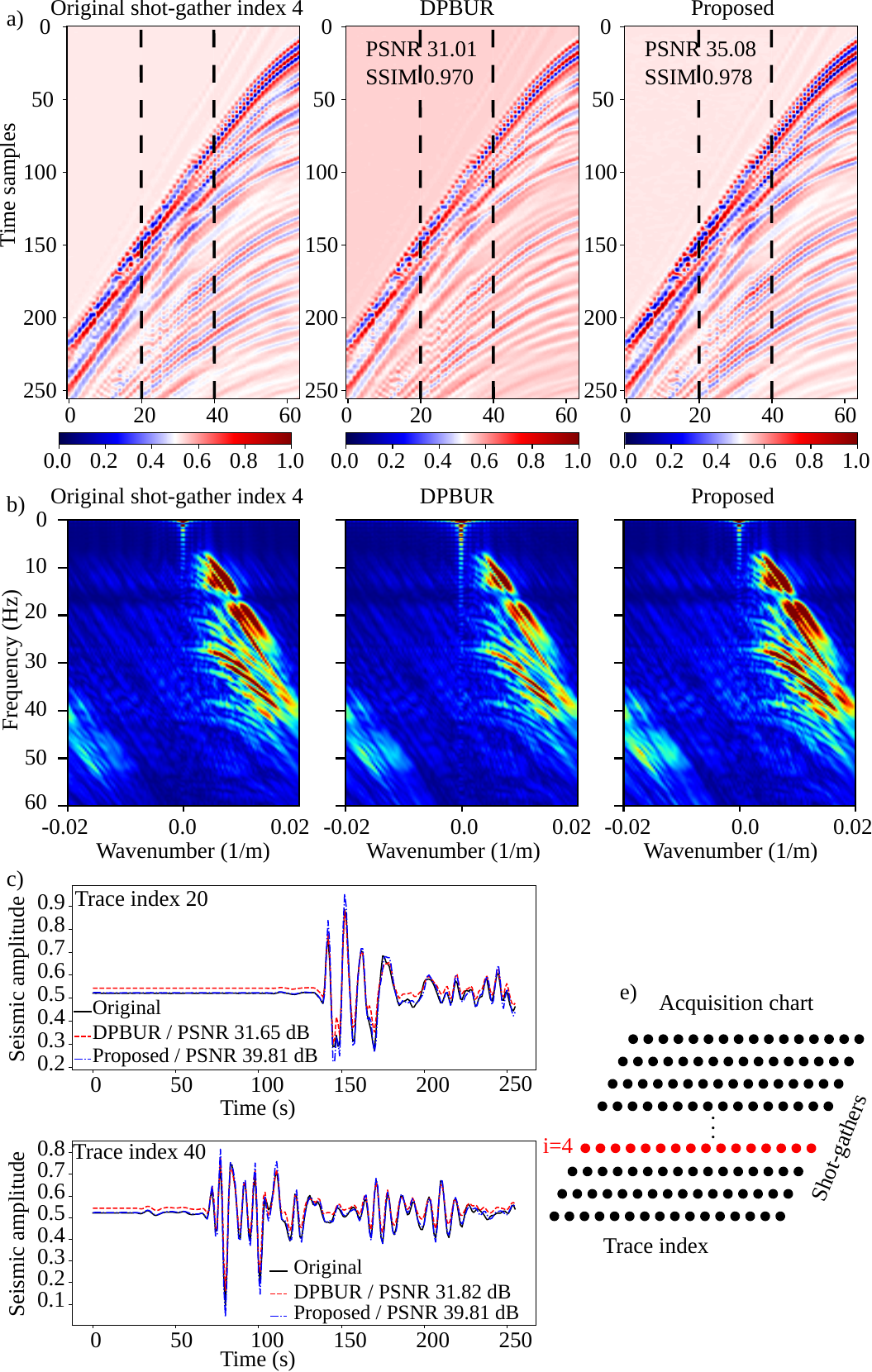}
    \caption{a) Reconstruction of shot-gather at index 4 from field Alaska dataset in Experiment III using DPBUR~\cite{Kong2022} and the proposed method. b)  Data in the frequency-wavenumber (FK) domain.  c) Trace comparison at indices 40 and 20 from shot-gather $4$ depicted as vertical dashed lines in a).  d) Acquisition chart displaying the missing shot-gather ($i=4$). Removing a shot-gather involves removing all traces associated with its corresponding active receivers (red dots). } 
    \label{fig:exp3}
\end{figure} 
 
We compared the reconstruction results from the proposed method with those obtained with the Deep prior-based unsupervised reconstruction (DPBUR) \cite{Kong2022}. It is worth noting that DPBUR was specifically designed to address the reconstruction of 3D seismic data. We simulated a scenario with $\sim 36\%$ missing shot-gathers, corresponding to the indices $\lbrace 4,8,10,12,15,17,21,23,24,26\rbrace$. 

Table ~\ref{table:exp3} summarizes the average reconstruction results for the recovered shot-gathers in terms of PSNR and SSIM metrics. These results demonstrate that our method exhibits a high generalization capacity, outperforming DPBUR in up to $6.2$ dB of PSNR and $0.013$ SSIM. The main differences in the results are due to amplitude overestimations from DPBUR, which in turn imply trace shiftings, as depicted in the traces comparison from Fig.~\ref{fig:exp3}c. Further, the recovered shot-gathers illustrated in Fig.~\ref{fig:exp3}a--b show that the proposed method provides a better estimation of typical seismic structures such as reflections (hyperbolic events) and refractions (linear events) \rev{both in signal and FK domains}. 

\begin{comment}
   \begin{figure}[ht]
    \centering
    \includegraphics[width=.8\columnwidth]{figures/exp3traces.pdf}
    \caption{Comparison of original and reconstructed traces from Experiment III a) Trace index 40 and b) Trace index 20 from shot-gather $4$. DPBUR overestimates the reconstructed seismic amplitudes.%, exhibiting a 2B lower quality compared to the proposed method.
    }
    \label{fig:exp3traces}
\end{figure} 
\end{comment}

\begin{table}[ht]
\centering
\caption{Average PSNR and SSIM for the reconstructed shot-gathers of Mobil AVO viking dataset in Experiment III. }
\label{table:exp3}

\begin{tabular}{ccc}
\hline
Method    & SSIM                   & PSNR (dB)               \\ \hline
Proposed & \textbf{0.985 $\pm$ 0.005} & \textbf{37.172 $\pm$ 1.149} \\ \hline
DPBUR~\cite{Kong2022}     & 0.972 $\pm$ 0.012          & 30.930 $\pm$ 0.950          \\ \hline
\end{tabular}
\end{table} 

\subsection{\claudia{Experiment IV}}
\begin{figure}[b!]
    \centering
    \includegraphics[width=.5\columnwidth]{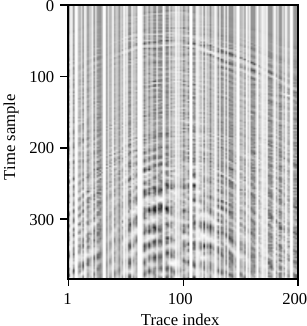}
    \caption{Example of  subsampled \textit{SEAM Phase II}/foothills shot with 40\% missing traces.}% in the \textit{SEAM Phase II}/foothills dataset}
    \label{fig:exp4_subsam}
\end{figure}

\claudia{In this experiment, we tested the proposed method in a scenario characterized by high noise levels and, seismic events at various inclination angles and frequencies. \claudia{Therefore, we used the well-known \textit{SEAM Phase II}/foothills dataset \cite{Regone2017}}. Additionally, we analyzed different compression rates (percentage of traces removed): 20\%, 30\%, 40\%, and 50\%. Compression rates exceeding 50\% were not considered as our algorithm is primarily designed to complement seismic preprocessing tasks downstream. Thus, up to 50\% missing traces suffice for noisy data. An example of a subsampling mask to remove 40\% traces is illustrated in Fig.~\ref{fig:exp4_subsam}. Results from this experiment are compared with two classical model-based methods, that address the problem %from the perspective of %algorithms model-based on 
exploiting signal sparsity and smoothness \cite{Ma2017,Baraniuk2017}.} % promotion.} 
%\rev{We compare the proposed solution with 2 classical approaches based on sparsity promotion \cite{Ma2017,Baraniuk2017} and smoothness priors as follows: }

\claudia{Specifically, the sparsity-promotion inversion exploits the signal sparseness in the Frequency-wavenumber (FK) domain via a $\ell_2-\ell_1$ minimization given by  $J = \|\mathbf{y} - \mathbf{R} \mathbf{F}^H \mathbf{x}\|_2 +\epsilon \|\mathbf{F}^H \mathbf{x}\|_1$, where $\mathbf{F}$  is the sparsity basis operator, $\mathbf{x}$ and $\mathbf{y}$ are the complete and subsampled signals, respectively, and $\mathbf{R}$ is the subsampling operator. The regularization parameter $\epsilon$ controls the sparsity strength, and for our experiments, its value was set to $\epsilon=0.1$. More details on this formulation for seismic interpolation problems can be found in \cite{Goyes-Penafiel2021, Baraniuk2017}.}

%\begin{itemize}
%    \item \rev{The sparsity-promotion inversion exploits the signal sparseness in the Frequency-wavenumber (FK) domain via a $\ell_1$ by minimizing  $J = \|\mathbf{y} - \mathbf{R} \mathbf{F}^H \mathbf{x}\|_2 +\epsilon \|\mathbf{F}^H \mathbf{x}\|_1$, where $\mathbf{F}$  is the transform function, $\mathbf{x}$ and $\mathbf{y}$ are the completed and subsampled signal, and $\mathbf{R}$ is the restriction operator modeling the acquisition positions. The parameter $ \epsilon$ controls the sparsity strength, for our experiments was set to $\epsilon=0.1$.}
%    \item 

\begin{figure*}[ht]
    \centering
    \includegraphics[width=.9\textwidth]{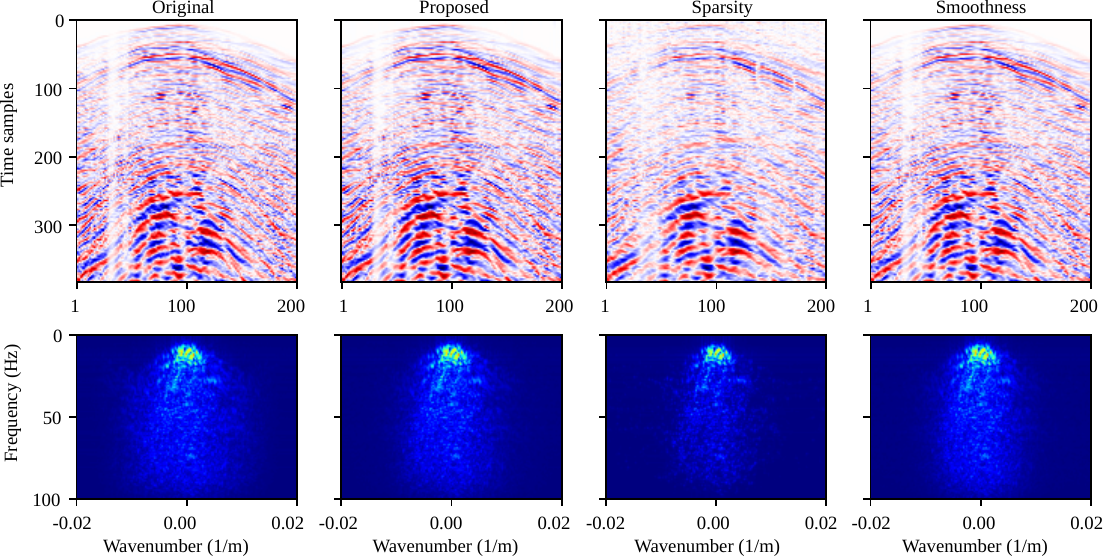}
    \caption{Visual reconstruction of subsampled data from Fig. \ref{fig:exp4_subsam} ($40\%$ missing traces) using sparsity and smoothness inversion methods compared with the proposed (deep learning-based) method. Data in the normalized frequency-wavenumber domain in the bottom row.}
    \label{fig:exp4_recons}
\end{figure*}

\claudia{On the other hand, the regularized inversion exploits the signal smoothness along the spatial axis, i.e., the second derivative operator $\nabla$, by minimizing $J = \|\mathbf{y} - \mathbf{R} \mathbf{x}\|_2 + \epsilon_\nabla ^2 \|\nabla \mathbf{x}\|_2$, where $\epsilon_\nabla$ controls the strength of the smoothness term. For our experiment, the value of $\epsilon_\nabla$ was set to $\sqrt{0.1}$.}
%\end{itemize}
\claudia{%We adopted the solution for 
Both inversion problems %sparsity-promotion and smoothness
were solved using the linear-operator library PyLops \cite{Ravasi2020PyLopsAOptimization}. }

\begin{table}[h!]
\centering
\caption{Average PSNR and SSIM for different compression ratios compared with classical interpolation methods.}
\label{table:exp4}
\begin{tabular}{clll}
\cline{2-4}
\multicolumn{1}{l}{} & Method   & PSNR (dB) & SSIM \\ \hline
                     & Proposed &   34.150        & \textbf{0.985}     \\ \cline{2-4} 
20\%                 & Sparsity &   27.103        & 0.853     \\ \cline{2-4} 
\multicolumn{1}{l}{} & Smooth   &   \textbf{35.670}        & \textbf{0.985}      \\ \hline \hline
                     & Proposed &   33.041         &  \textbf{0.978}    \\ \cline{2-4} 
30\%                 & Sparsity &   26.161         &  0.815    \\ \cline{2-4} 
\multicolumn{1}{l}{} & Smooth   &   \textbf{33.653}         &  0.976    \\ \hline \hline
                     & Proposed &   \textbf{31.361}         &  \textbf{0.966}     \\ \cline{2-4} 
40\%                 & Sparsity &   24.855         &  0.758    \\ \cline{2-4} 
\multicolumn{1}{l}{} & Smooth   &   31.201         &  0.960    \\ \hline \hline
                     & Proposed &   \textbf{27.759}         &  \textbf{0.927}     \\ \cline{2-4} 
50\%                 & Sparsity &   23.394         &  0.662    \\ \cline{2-4} 
\multicolumn{1}{l}{} & Smooth   &   27.362         &  0.897    \\ \hline
\end{tabular}
\end{table}

\claudia{Fig~\ref{fig:exp4_recons} illustrates the visual reconstruction results for $40\%$ missing traces, which were randomly and irregularly removed as depicted in Fig. \ref{fig:exp4_subsam}. It can be seen that the proposed method and smoothness adequately reconstructed the seismic events, preserving the frequency contents of the signal, while the sparsity method introduces artifacts due to the high noise content. %which causes the signal to be not completely sparse in the FK domain.
The PSNR and SSIM metrics for different compression rates are summarized in Table~\ref{table:exp4}, where it can be noted that the proposed method outperforms the classical methods for \claudia{higher compression rates (40\% and 50\%), while exhibiting comparable results on lower compression scenarios, where the inversion problem is less complex.}}

\section{Discussion}

\claudia{ Section \ref{sec:experiments} tested the capabilities of the proposed deep-learning-based approach for seismic data reconstruction in multiple scenarios, including field and synthetic datasets. These scenarios encompassed various levels of noise, as well as seismic events with different characteristics. A common aspect of all the experiments is that network training was based solely on synthetic data, reducing dependencies on large real data sets. The reconstruction results demonstrate that, for the studied scenarios, synthetic data suffices to model seismic features of different complexities. Despite these promising results, we believe there is still room for improvement by enhancing the method's generalization capacity through the inclusion of real data during network training. This would allow a wider variety of seismic features to be considered within the model, thereby achieving more robust results.
}

\section{Conclusions}

A GAN-supervised learning-based method for seismic data reconstruction was introduced. The proposed method addresses the persistent domain generalization challenge of existing reconstruction approaches. In contrast to conventional methods that heavily rely on extensive training datasets or employ techniques like transfer learning to enhance generalization, our approach strategically introduces controlled data generation during network training. This not only mitigates overfitting but also substantially enhances domain generalization without using any external field data in the reconstruction model.

The experimental results for trace reconstruction from pre-stack and post-stack data, and shot-gather reconstruction from a marine 2D survey show the reliable performance of our proposed approach compared to both traditional supervised learning methods and unsupervised techniques like Deep Image Prior and Internal Learning. Quantitatively, the GAN-supervised approach consistently outperforms traditional supervised learning by 6.6 dB and 1.4 dB in PSNR for synthetic and field data, respectively, and exhibits up to 8 dB improvements in PSNR over unsupervised learning in post-stack data reconstruction. Furthermore, our method excels in shot-gather reconstruction, surpassing Deep Prior-based Unsupervised Reconstruction (DPBUR) by 6.2 dB in PSNR and 0.013 in SSIM. This method has shown a great capability to enhance data while promoting accurate seismic images in downstream processing flows.

\section*{Acknowledgment}
The authors thank the  NVIDIA Academic Hardware Grant Program and the High-Dimensional Signal Processing (HDSP) lab for providing computational resources to run the experiments.

\section*{Data and Materials Availability}
For the sake of reproducibility, data, and codes to replicate the described experiments are accessible at  \url{https://github.com/PAULGOYES/GAN_guided_seismic}. 
\ifCLASSOPTIONcaptionsoff
  \newpage
\fi

\bibliographystyle{IEEEtran}
\bibliography{mendeley,others}

\newpage

\section{Biography Section}
%If you have an EPS/PDF photo (graphicx package needed), extra braces are needed around the contents of the optional argument to biography to prevent the LaTeX parser from getting confused when it sees the complicated $\backslash${\tt{includegraphics}} command within an optional argument. (You can create your own custom macro containing the $\backslash${\tt{includegraphics}} command to make things simpler here.)
 %#\vspace{11pt}

%\bf{If you include a photo:}\vspace{-33pt}

\begin{comment}
    \begin{IEEEbiography}[{\includegraphics[width=1in,height=1.25in,clip,keepaspectratio]{photos/fig1.png}}]{Michael Shell}
Use $\backslash${\tt{begin\{IEEEbiography\}}} and then for the 1st argument use $\backslash${\tt{includegraphics}} to declare and link the author photo.
Use the author name as the 3rd argument followed by the biography text.
\end{IEEEbiography}
\end{comment}


\begin{IEEEbiographynophoto}{Paul Goyes-Pe\~nafiel} (Graduate Student Member, IEEE)
received a B.Sc. in Geology from the Universidad Industrial de Santander, Bucaramanga, Colombia, and M.Sc. in Geophysics from the Perm State University, Perm, Russia, in 2009 and 2018 respectively. He is currently a Ph.D. candidate in Computer Science with the Universidad Industrial de Santander. He was a research intern (Sept 2023 to Dec 2023) at Washington University in St. Louis. Paul has been a researcher in the hydrocarbon industry, specializing in applied geophysics for shallow and deep exploration. His research interests are inverse theory, seismic acquisition, and processing, electromagnetic techniques for hydrogeological and crustal exploration, as well as the integration of advanced deep learning methodologies in geosciences.
\end{IEEEbiographynophoto}

\begin{IEEEbiographynophoto}{León Suárez-Rodríguez}
    received the B.S.E. degree in Civil Engineering (2023) from Universidad Industrial de Santander, Colombia. He is currently pursuing his Master’s degree in Systems Engineering at the same university. His research interests focus on inverse problems and deep learning applications in geosciences and computational imaging.
\end{IEEEbiographynophoto}

\begin{IEEEbiographynophoto}{Claudia V.  Correa}
(Member, IEEE) received the B.Sc. and M.Sc. degrees in Computer Science from the Universidad Industrial de Santander, Bucaramanga, Colombia, in 2009 and 2013, respectively, and the M.Sc. and Ph.D. degrees in electrical and computer engineering from the University of Delaware, Newark, DE, USA, in 2013 and 2017, respectively. She is currently a research fellow at the Computer Science Department, Universidad Industrial de Santander. Her research interests include computational imaging, compressive spectral imaging, and machine learning applications to seismic problems.
\end{IEEEbiographynophoto}

\begin{IEEEbiographynophoto}{Henry Arguello}
(Senior Member, IEEE) received the B.Sc. Eng. degree in electrical engineering and the M.Sc. degree in electrical power from the Universidad Industrial de Santander, Bucaramanga, Colombia, in 2000 and 2003, respectively, and the Ph.D. degree in electrical engineering from the University of Delaware, Newark, DE, USA, in 2013. He is currently an Associate Professor with the Department of Systems Engineering, Universidad Industrial de Santander, and associate editor for IEEE Transactions on Computational Imaging, and the president of the signal processing chapter in Colombia. In 2020, he was a Visiting Professor at Stanford University, Stanford, CA, USA, funded by Fulbright. His research interests include high-dimensional signal processing, optical imaging, compressed sensing, hyperspectral imaging, and computational imaging.
\end{IEEEbiographynophoto}

\vfill

\end{document}